\documentclass{aa}
\usepackage{txfonts}
\usepackage{graphicx}
\usepackage{lscape}
\usepackage{xcolor}

\usepackage{natbib}
\bibpunct{(}{)}{;}{a}{}{,}

\usepackage[normalem]{ulem}

\usepackage{pdflscape}
\usepackage{subfigure}
\usepackage{array}
\usepackage{multirow}
\usepackage{multicol}
\usepackage{longtable}
\usepackage{supertabular}

\usepackage{graphicx}
\usepackage{txfonts}

\usepackage{hyperref}
\begin{document}

\title{Spectroscopic detection of a 2.9-hour orbit in a Long Period Radio Transient}

\author{Antonio C. Rodriguez\thanks{Email: acrodrig@caltech.edu}}

\institute{Department of Astronomy, California Institute of Technology, 1200 E California Blvd, Pasadena, CA, 91125, USA}

\date{}


  \abstract{Long Period radio Transients (LPTs) are a mysterious new class of radio transients pulsating on periods of minutes to hours. So far, nine LPTs have been discovered predominantly at low Galactic latitudes, yet their nature remains unknown. Here, I present the first phase-resolved optical spectroscopy of the 2.9-h LPT GLEAM-X J0704--37, acquired with the 10-m Keck I telescope. Radial velocity (RV) shifts of $189\pm 3\;\textrm{km s}^{-1}$ of an M5-type star in a binary system are detected on a period nearly equal to the radio period. Weak H$\alpha$ emission is also present, with some of it possibly originating from outside of the M dwarf. Based on the RV amplitude, and assuming a typical M dwarf mass, the companion mass must be $M \geq 0.22 M_\odot$. Calibrating the spectra with space-based \textit{Gaia} photometry reveals that the system is nearly four times closer than previously reported, at $d \approx 400$ pc, suggesting that more systems could be nearby and amenable to optical characterization. The optical spectrum between 3500--10,000 \AA\;is well modeled by a binary comprised of a massive white dwarf (WD; $T_\textrm{eff}\approx$7,300 K, $M\approx0.8-1.0M_\odot$) and M dwarf ($T_\textrm{eff}\approx$3,000 K, $M\approx0.14M_\odot$). Radio pulses arrive when the WD is at nearly maximum blueshift and the M dwarf at nearly maximum redshift, in contrast to what has been reported in a similar LPT, ILT J1101+5521. GLEAM-X J0704--37 is now the second LPT with an orbital period nearly equal to the radio period, establishing a class of LPTs associated with WD + M dwarf binaries; other LPTs are likely related related to WD and/or neutron star spins. This work demonstrates that precise localization of LPTs, which enables optical follow-up, will be key in uncovering the mechanism(s) that power this new class of phenomenon.}

   \maketitle
%

\section{Introduction}
Long period radio transients (LPTs) represent a new era in radio astronomy. Pulsars were the first radio sources discovered to be stably pulsating at $\sim$1 sec timescales \citep{1968hewish}, and attributed to rotating neutron stars \citep[NSs;][]{1968gold, 1968pacini}. Exploring a new range of periodicity has historically proven to be revolutionary in the field. For instance, the discovery a millisecond pulsar \citep[MSP;][]{1982backer} and subsequent studies established the importance of binary interaction in the formation of such rapid rotators \citep[e.g.][]{1994pulsars}. 

Crucially, the association of optical (and other multiwavelength) counterparts with radio sources has proven to be vital for their characterization. \cite{1986kulkarni} optically identified WD binary companions to MSPs, establishing both the old age of MSPs and demonstrating that NS magnetic fields do not strictly decay exponentially. In addition, optical spectroscopy of the faint companions to black widow MSPs \citep{1988millisecond} with 10-m class telescopes has revealed what may be the most massive NS known to date, placing useful constraints on the NS equation of state \citep[$\approx2.35M_\odot$;][]{2022romani}. 

At the time of writing, nine LPTs have been reported\footnote{PSR J0901-4046 is an ultra-long period pulsar (not considered here to be an LPT) which has been established to be a neutron star with a 76 s period \citep{2022caleb}.} (in published form or as a preprint), and are summarized in Table \ref{tab:list}. \cite{2024discovery} reported the discovery of the 2.915-h LPT, GLEAM-X J0704--37. The radio source was localized using MeerKAT and associated with a faint (\textit{Gaia} G $\approx20.78$) optical counterpart at that position. An optical spectrum in that work revealed an M dwarf counterpart, similar to the one seen in the LPT ILT J1101+5521 \citep{2024deruiter}. \cite{2024discovery} also found a $\approx$6 yr period in radio timing residuals, and suggested it to be the binary period of a WD + M dwarf system, with the 2.9 h period being the WD spin. 

\begin{table*}[]
    \centering
    \begin{tabular}{l|c|c|c}
         LPT Name & Radio Period (min) & Counterpart? & Reference \\\hline
GCRT J1745-3009 & 76.2 & None & \cite{2005hyman} \\
GLEAM-X J162759.5-523504.3 & 18.18 & None & \cite{2022hurley-walker} \\
GPM J1839-10 & 21.97 & None & \cite{2023hurley-walker} \\
ASKAP J193505.1+214841.0 & 53.76 & NIR source & \cite{2024caleb} \\
CHIME J0630+25 & 7.017 & None & \cite{2024dong} \\
ILT J1101+5521 & 125.5& M dwarf & \cite{2024deruiter} \\
GLEAM-X J0704--37 & 174.9 & M dwarf & \cite{2024discovery} \\
ASKAP/DART J1832-0911 & 44.27 & X-ray source & \cite{2024wang, 2024li} \\
ASKAP J183950.5-075635.0 & 387 & None & \cite{2025lee}
    \end{tabular}
    \caption{List of the nine known LPTs at the time of writing (reported in published form or as a preprint).}
    \label{tab:list}
\end{table*}

Here, I show that GLEAM-X J0704--37 is in fact a \textit{compact} WD + M dwarf binary, with the 2.9 h radio period matching the orbital period and located only $\approx$400 pc away. Based on the orbital solution, radio pulses arrive when the WD is nearly at maximal blueshift and M dwarf nearly at maximal redshift.

\section{Data}
\label{sec:data}
GLEAM-X J0704--37 was observed on two occasions using the Low Resolution Imaging Spectrometer \citep[LRIS;][]{lris} on the 10-m Keck I telescope on Mauna Kea in Hawai'i. An observing log is provided in Appendix Table \ref{tab:data}. All LRIS data were wavelength calibrated with internal lamps, flat fielded, and cleaned for cosmic rays using \texttt{lpipe}, a pipeline for LRIS optimized for long slit spectroscopy \citep{2019perley_lpipe}. To calculate radial velocities (RVs), the mid-exposure time of all observations was corrected to the barycentric Julian date (BJD$_\textrm{TDB}$).

The spectrum shown in Figure 3 of \cite{2024discovery}, at  $\lambda\approx6000\AA$, ($F_\lambda\approx 3\times 10^{-18}\textrm{erg s}^{-1}\textrm{cm}^{-2}{\AA}^{-1}$) appears to be nearly four times lower that what is derived from the average \textit{Gaia} G photometry, $F_\lambda\approx 1.5\times 10^{-17}\textrm{erg s}^{-1}\textrm{cm}^{-2}{\AA}^{-1}$. Here, \textit{Gaia} Data Release 3 \citep[DR3;][]{2023gaiadr3} average photometry of the source (Gaia DR3 5566254014771398912) was used to calibrate the overall flux level obtained by spectroscopy. The \texttt{pyphot} package\footnote{\url{https://mfouesneau.github.io/pyphot/index.html}} was used to calculate synthetic photometry from the average spectrum of Night 2 in the \textit{Gaia} BP, G, and RP bandpasses. The average spectrum was then multiplied by a correction factor in order to match the synthetic photometry to the \textit{Gaia} photometry, leading to a smaller inferred distance ($\approx$400 pc) compared to that presented in the discovery paper ($\approx$1,500 pc)\footnote{Curiously, the nearer distance agrees with that derived from the radio dispersion measure using the electron density model of \cite{2017yao}, as reported in the discovery paper.}.

\section{Analysis}
\label{sec:analysis}
\subsection{RV fitting and mass constraints on companion star}

The radial velocity (RV) measurements traced by the Na I doublet (8183 and 8195 $\AA$) absorption lines on the left panel are shown in Figure \ref{fig:rv}. The full sets of spectra are shown in Appendix Figure \ref{fig:spectra}, and all RV measurements are reported in Appendix Table \ref{tab:rv}. The orbital phase is set to coincide with the start of observations on Night 2. Gray vertical lines indicate the observed radio pulses (2018-02-04 at 14:57:02, 2023-10-04 at 02:33:02, and 2023-10-04 at 05:29:59 UT) reported by \cite{2024discovery}, folded on the 10496 s period, and barycentric-corrected to BJD$_\textrm{TDB}$. Orbital period constraints of $10496\pm5$ s ($10500\pm500$ s) based on RVs from two (one) nights of data are outlined in Appendix Section \ref{sec:appendix_orbit}.

\begin{figure*}
    \centering
    \includegraphics[width=0.6\textwidth]{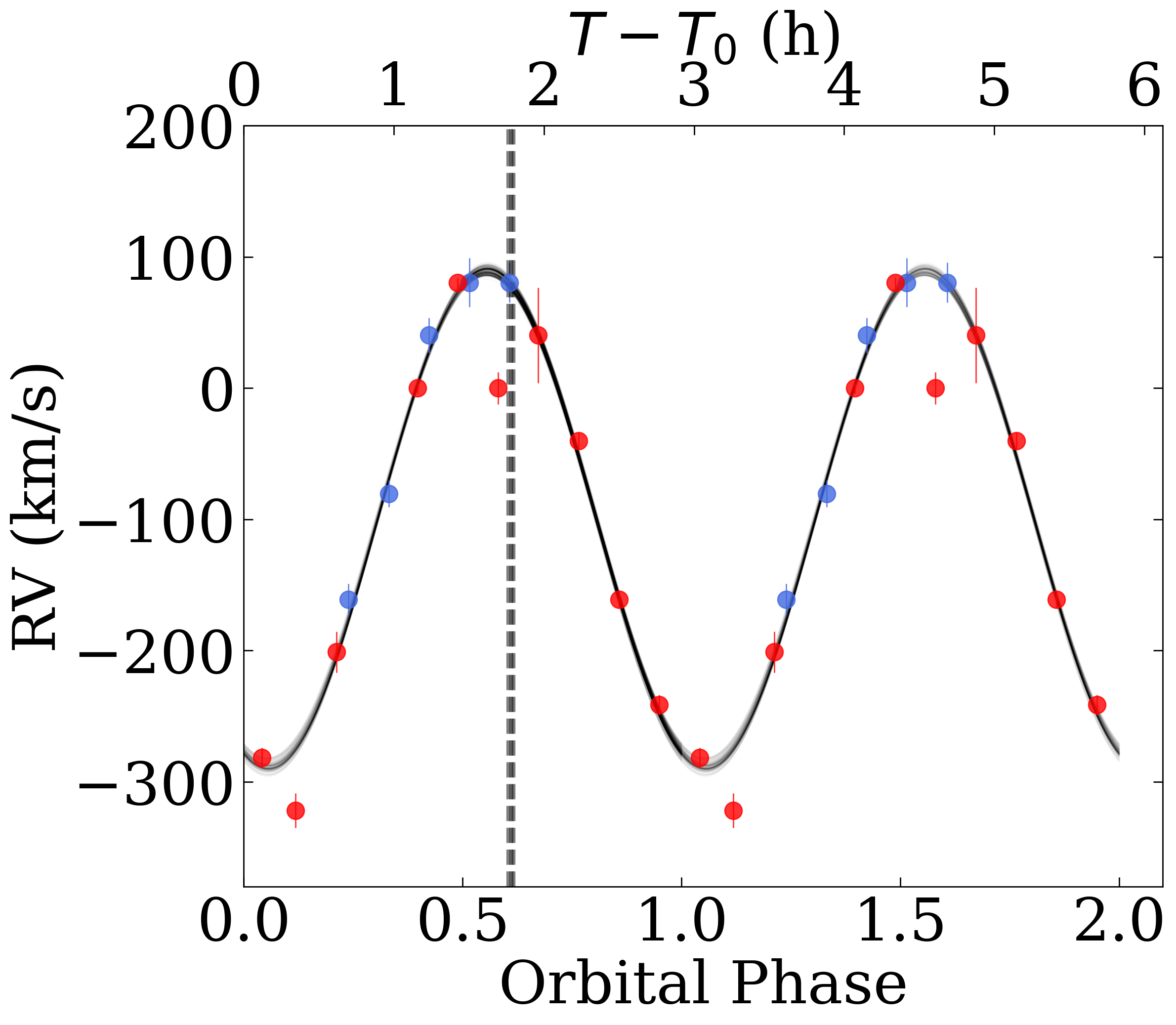}\includegraphics[width=0.3\textwidth]{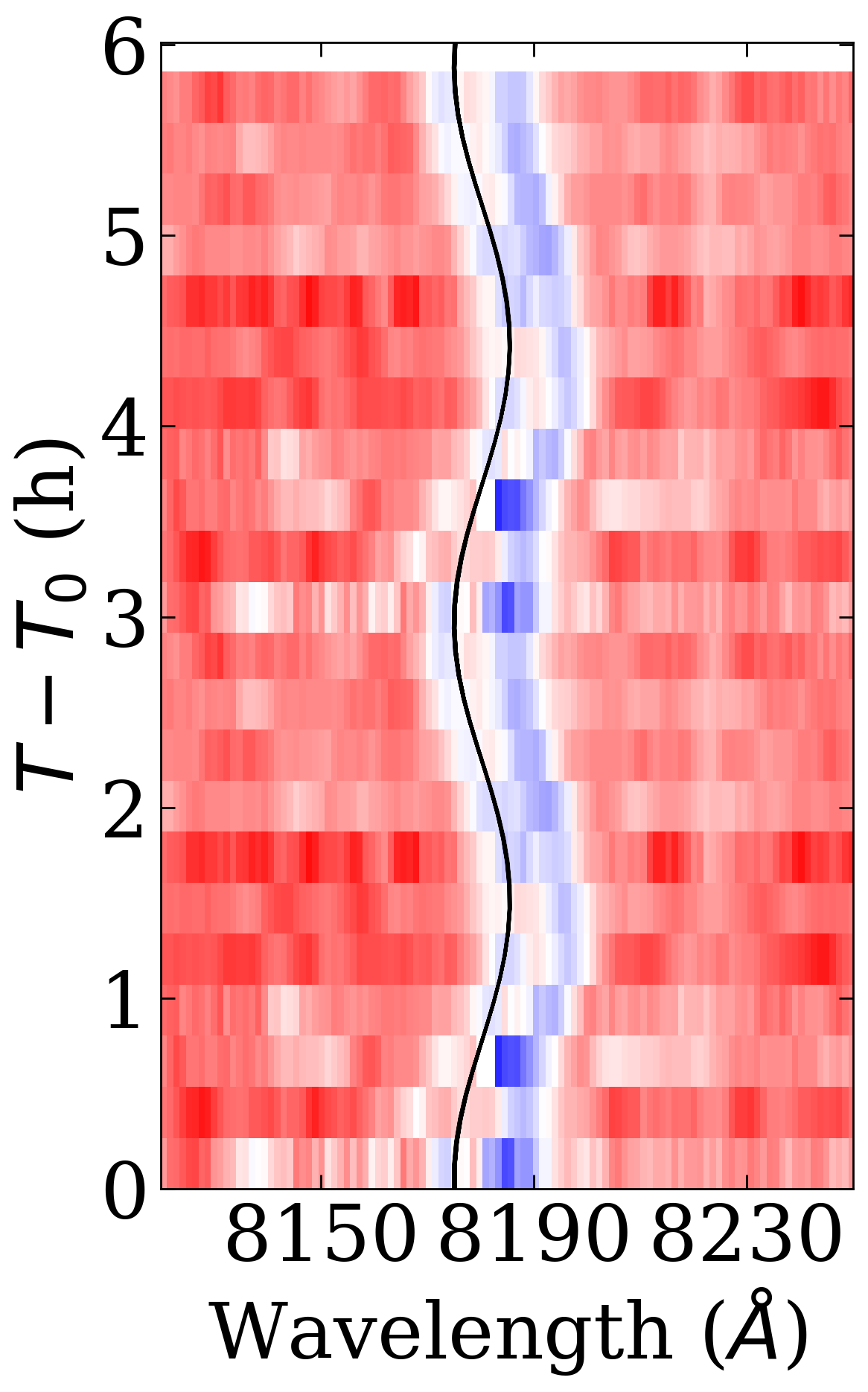}
    \caption{\textit{Left:} RVs measured using the Na I absorption doublet from Night 1 (blue) and Night 2 (red) are folded on a period of 2.915(1) h and plotted with the best RV model (black; $K_\textrm{MD}=189\pm3\;\textrm{km s}^{-1}$), demonstrating that the binary orbital period matches the radio pulse period to within 0.05 percent. Radio pulses (gray) occur just after the maximal redshift of the M dwarf, when it is at the ascending node as viewed from Earth. \textit{Right:} Trailed spectra (Night 2) of the Na I doublet show good agreement with the RV model (black). Two orbital periods are shown for visual aid.
    }
    \label{fig:rv}
\end{figure*}

Equation \ref{eq:mass_binary} yields a minimum mass constraint on the companion star:
\begin{equation}
    \frac{(M_\textrm{companion}\sin i)^3}{(M_\textrm{companion} + M_\textrm{MD})^2} = \frac{P_\textrm{orb}K_\textrm{MD}^3}{2\pi G}
        \label{eq:mass_binary}
\end{equation}
Assuming an edge-on inclination ($i = 90^\circ$) and an M dwarf of zero mass, the minimum companion mass is $M_\textrm{companion}\geq0.087 M_\odot$. Assuming a realistic M dwarf mass of $0.14M_\odot$, based on the spectral analysis below, yields $M_\textrm{companion}\geq0.22M_\odot$\footnote{\textrm{It has been found, that in M dwarfs in close binaries, the Na I doublet may trace the back of the star, leading to the need for a correction to the RV amplitude of at most $\approx -15\;\textrm{km s}^{-1}$ \citep[e.g.][]{1988wade}. This would imply a minimum companion mass of 0.068$M_\odot$ and 0.20$M_\odot$, assuming an M dwarf mass of zero and 0.14$M_\odot$, respectively.}}.

\subsection{H$\alpha$ emission associated with the M dwarf}
\label{sec:halpha}
\begin{figure}
    \centering
    \includegraphics[width=0.3\textwidth]{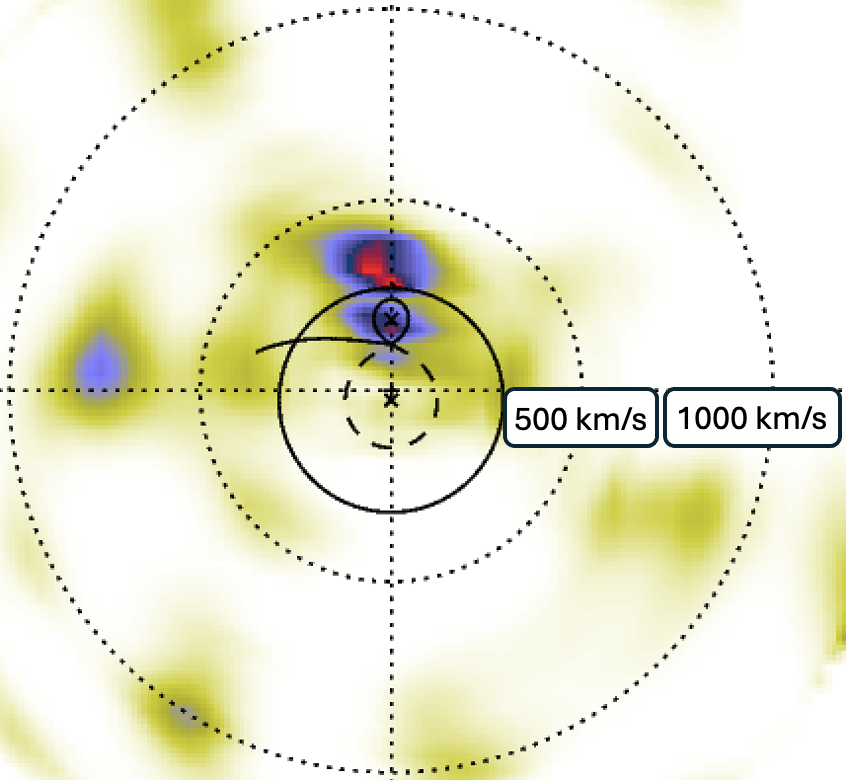}\includegraphics[width=0.19\textwidth]{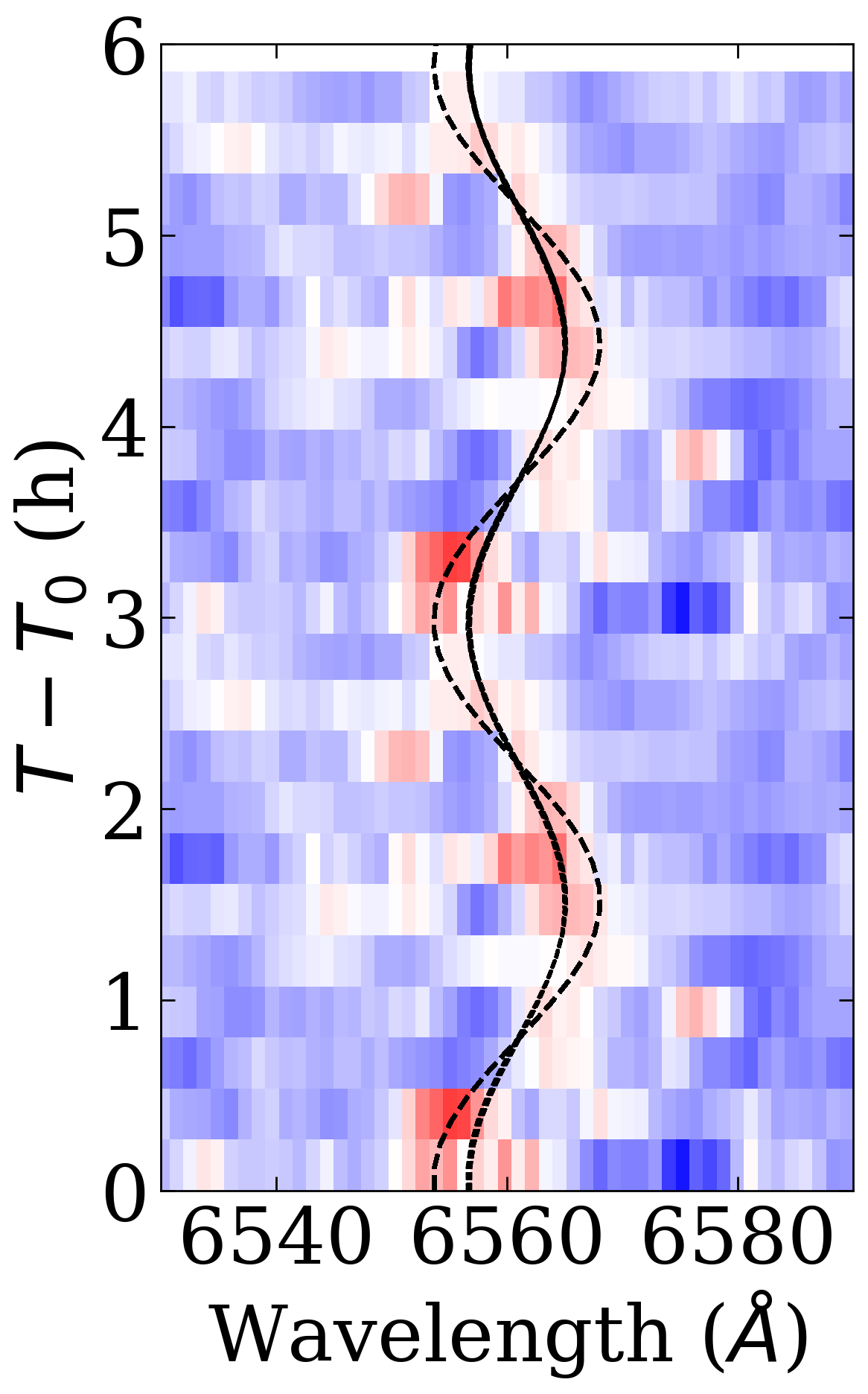}
    \caption{\textit{Left:} Doppler tomogram of the H$\alpha$ line reveals $>3\sigma$ (red) emission associated with the orbital phase of the M dwarf (Roche potential shown by the solid black teardrop shape). The majority of the emission may originate from \textit{outside} the M dwarf, with a higher observed velocity. \textit{Right:} Trailed spectra of the H$\alpha$ emission line shown alongside the M dwarf RV model (black solid line), and the higher velocity component (black dotted line; $K\approx330\;\textrm{km s}^{-1}$).}
    \label{fig:map}
\end{figure}

The origin of the H$\alpha$ emission is explored by creating a ``Doppler tomogram''\footnote{A Doppler tomogram, in polar coordinates, shows RV shift as the radial coordinate and orbital phase as the azimuthal coordinate. Doppler maps are commonly used in the cataclysmic variable (CV) literature to disentangle emission from the donor star and accretion-related components, though they are also useful for detached binaries; see \cite{2001marsh} for a review of the method of Doppler tomography.} \citep{1988marsh} in Figure \ref{fig:map} with the \texttt{doptomog} IDL-based code\footnote{\url{https://www.saao.ac.za/~ejk/doptomog/main.html}} \citep{2015doptomog}. The systemic offset of $\gamma=-98.9\;\textrm{km s}^{-1}$ was applied, and orbital phases were shifted such that inferior conjunction corresponds to phase of $90^\circ$ on the resulting map \citep[e.g.][]{2015doptomog, 2023rodriguez}. The Roche potentials derived from the inferred masses and system inclination (Table \ref{tab:params}) are overlaid, showing the Roche potential of the M dwarf as a solid black teardrop\footnote{Typically, in the CV literature, this also represents the actual shape of the star since it fills its Roche lobe in such systems. That is, however, not the case here since the M dwarf in GLEAM-X J0704--37 does not fill its Roche lobe.}. The dotted black teardrop is the Roche potential of the companion star. Other markings are related to accreting systems and can be ignored here. The physical geometry mapped onto these coordinates is illustrated in Figures 1 and 3 of \cite{2015doptomog}.

Figure \ref{fig:map} shows that some H$\alpha$ emission is coincident with the orbital phase and RV of the donor star (inside the Roche potential), but that the bulk of the H$\alpha$ emission may originate from a higher velocity region, outside of the M dwarf. However, this is only seen in the single orbit of spectra presented here, so further data is needed to confirm the presence of this feature. It may very well turn out that all H$\alpha$ emission is associated with chromospheric activity of the rapidly rotating M dwarf \citep[e.g.][]{1998delfosse}. If this higher velocity feature is indeed present, then its velocity corresponds to $\approx 330\;\textrm{km s}^{-1}$, as represented by the black dashed curve in the right panel of Figure \ref{fig:map}. Details regarding the possible spatial origin of that emission region and resemblance to the post common envelope WD + M dwarf binary QS Vir are discussed in Appendix Section \ref{sec:halpha}. 

\subsection{Binary parameters}
\label{sec:parameters}

The spectra of GLEAM-X J0704--37 do not reveal any strong emission lines which could be indicative of accretion, suggesting this system is a detached binary (M dwarf + companion). Because there is a detection of a blue continuum (shortward of 5500 \AA) above the noise, I proceed with the assumption that the companion star is a white dwarf (WD), also adopted by \cite{2024discovery} based on radio pulse arguments. Further arguments against a NS + M dwarf binary are put forth in Appendix Section \ref{sec:appendix_wdns}.

Proceeding with this assumption, the average spectrum can be described by six parameters, only four of which are independent: the WD effective temperature and radius, $T_\textrm{WD}$ and $R_\textrm{WD}$; the M dwarf effective temperature and radius, $T_\textrm{MD}$ and $R_\textrm{MD}$; the distance to the system, $d$; and the reddening to the system, $E(B-V)$. $T_\textrm{MD}$ and $R_\textrm{MD}$ are related though the M dwarf isochrones (assuming an age of 10 Gyr) of \cite{2015baraffe}; $d$ and $E(B-V)$ are related through the 3D dust map of \cite{2024edenhofer}. Other dust maps infer slightly different levels of extinction, explored further in Appendix Section \ref{sec:mcmc_appendix}. The four independent parameters were constrained using a Markov Chain Monte Carlo (MCMC) parameter exploration, described in complete detail in Appendix Section \ref{sec:appendix_binary}. The resulting parameters are credible intervals are shown in Table \ref{tab:params}.

\begin{table}[]
    \centering
    \begin{tabular}{l|c}
        Parameter & Value \\
        \hline
        $K_\textrm{MD}$ (km s$^{-1}$)& $189.4 \pm 2.7$\\
        $\phi_0$ & $1.3074 \pm 0.0021$\\
        $\gamma$ (km s$^{-1}$) & $-98.9^{+1.7}_{-1.8}$\\
        $P_\textrm{orb}$ (s) & $10496 \pm 5$\\
         \hline
         $E(B-V)$ & $0.119\pm0.001$\\
         $T_\textrm{WD}$ (K) & $7320^{+800}_{-590}$\\
         $R_\textrm{WD}$ ($R_\odot$) & $0.0079^{+0.0014}_{-0.0015}$\\
         $T_\textrm{MD}$ (K) & $3010 \pm 20$\\
         $R_\textrm{MD}$ ($R_\odot$) & $0.165^{+0.003}_{-0.004}$\\
         $d$ (pc) & $380\pm10$\\
         \hline
         $M_\textrm{WD}$ ($M_\odot$) & $1.02^{+0.12}_{-0.13}$\\
         $M_\textrm{MD}$ ($M_\odot$) & $0.136 \pm 0.003$\\
         $a$ ($R_\odot$) & $1.07\pm0.04$\\
         $q$ & $0.133^{+0.019}_{-0.014}$\\
         $R_\textrm{MD}/R_L$ & $0.680 \pm 0.001$\\
         $i$ ($^\circ$) & $28^{+2}_{-1}$\\
    \end{tabular}
    \caption{All parameters for the GLEAM-X J0704--37 WD + MD binary system. The first level of parameters are just based on the RV shifts of the MD, while the next level is based on an MCMC parameter exploration of the average spectrum. The final level represents parameters derived from those used in the MCMC analysis.}
    \label{tab:params}
\end{table}

\begin{figure*}
    \centering
    \includegraphics[width=0.8\textwidth]{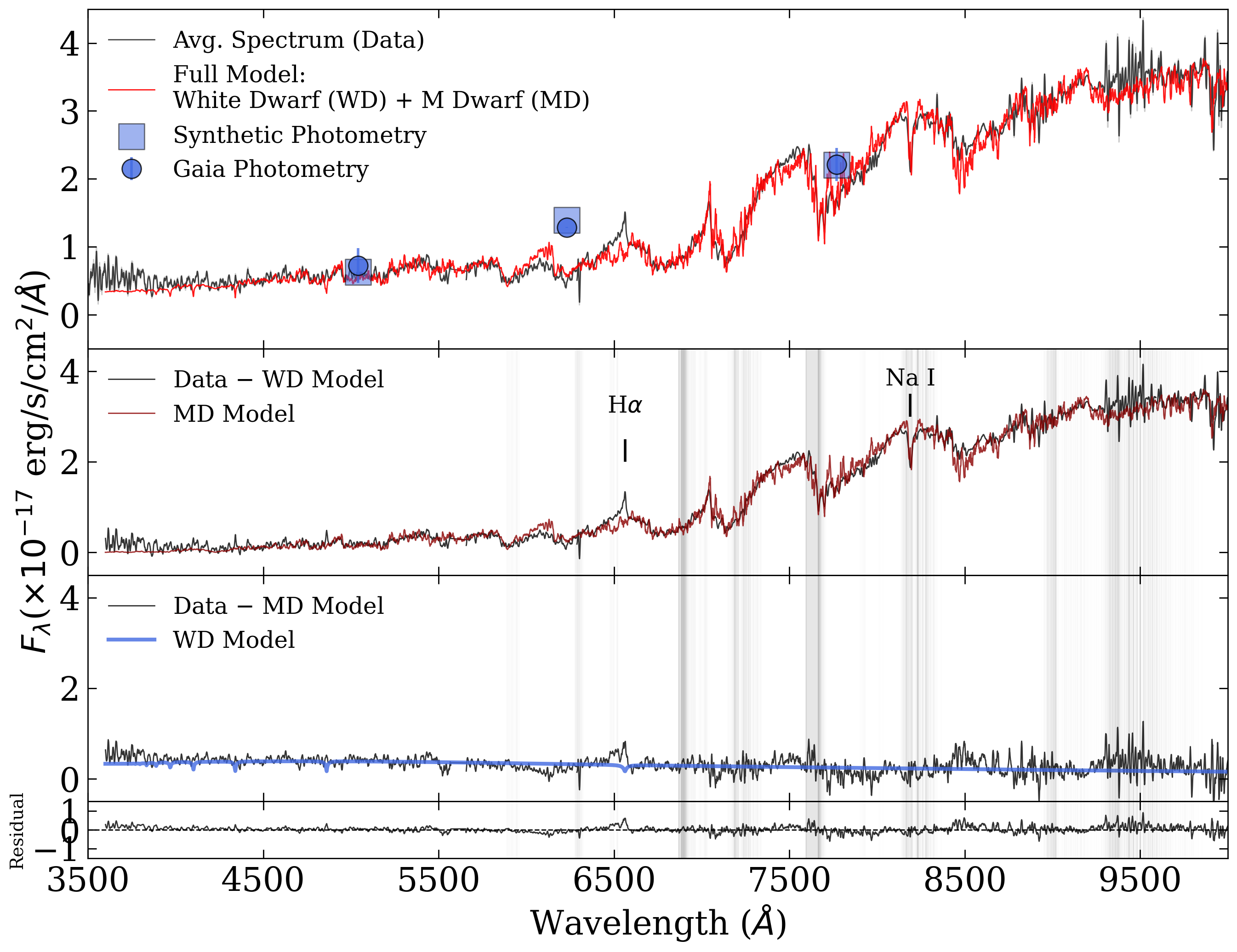}
    \caption{\textit{Top:} The average spectrum of GLEAM-X J0704--37 (black) is well fit by a WD + MD binary model (red; parameters in Table \ref{tab:params}). The overall flux level was calibrated by matching synthetic photometry (squares), generated from the spectrum, with \textit{Gaia} average photometry (circles). \textit{Middle:} Each binary component provides a good fit to the data when the other is subtracted. \textit{Bottom:} Residuals show no obvious systematic disagreement, aside from the H$\alpha$ emission and telluric features (gray bands).}
    \label{fig:best}
\end{figure*}

In Figure \ref{fig:best}, the average spectrum (rectified using \textit{Gaia} photometry), both \textit{Gaia} and synthetic photometry, and the model obtained from the parameters in Table \ref{tab:params} are shown. The residuals do not show any obvious systematic biases across the entire 3500--10,000\AA\; range, aside from features related to telluric subtraction and the H$\alpha$ emission feature at 6563\AA.

The largest uncertainty in the above analysis is the extinction to the system; assuming higher extinction values leads to a larger WD radius (meaning lower mass). The WD in this system, however, is clearly massive, since it is unlikely for the extinction to exceed $E(B-V) = 0.3$ (see Appendix \ref{sec:mcmc_appendix}). Assuming this leads to a WD mass of 0.82$M_\odot$, lower than that presented in Table \ref{tab:params}, but still more massive compared to typical single WDs in the field \citep[$0.6M_\odot$, $0.7M_\odot$ for DA and DB WDs;][]{2007wd} or \textrm{WDs in post common envelope binaries \citep[$0.67\pm0.21 M_\odot$; ][]{2011zorotovic}}.

\section{Discussion}
\label{sec:discussion}
\subsection{A detached WD + M dwarf close binary}

I have shown that the optical spectrum of GLEAM-X J0704--37 is well modeled by that of a detached (non-accreting) WD + MD binary with an orbital period of 10496$\pm$5 s (2.915 h), which agrees with the radio period (10496.5522) to within 0.05 percent; 5 percent at worst if only fitting RVs from Night 2. This discards a scenario put forth by \cite{2024discovery}, where the 2.9 h period was attributed to NS or WD spin period in a $\approx$ 6 yr orbital period with an M dwarf \footnote{A faint third body could still be present in the system orbiting on a 6 yr period. Recent discoveries in close compact object binaries have revealed the presence and evolutionary impact of distant triples \citep{2024burdge, 2024shariat}. At the same time, $\sim$yr long timing residuals seen in CVs have been attributed to magnetic activity in the low-mass companion star \citep{2024souza}. Both scenarios warrant further investigation to explain the radio timing residuals.}.

\subsection{Connections to ILT J1101+5521, WD pulsars, and cataclysmic variables}

GLEAM-X J0704--37 bears a strong similarly to another LPT, ILT J1101+5521 \citep{2024deruiter}. They both host cool, massive WDs ($\approx$7,300 K and $\approx$5,500 K;  $\approx1.02M_\odot$ and $\approx0.77M_\odot$, respectively). They are both WD + M dwarf binaries, pulsing in the radio on a similar nearly equal to the orbit. WD ``pulsars'', which are also radio emitting detached WD + M dwarf systems, but host a magnetic WD spinning and radio pulsing on 2--5 min timescales, though in a 3.5--4 h orbit, are another class of similar radio sources, though over an order of magnitude less radio luminous than LPTs \cite{2016marsh, 2022pelisoli}. All of these systems highlight the unexpected diversity of radio activity and emission processes associated with otherwise pedestrian detached WD + M dwarf binaries. These systems, including GLEAM-X J0704--37, will eventually begin mass transfer and become cataclysmic variables \citep[CVs;][]{1995hellier, 2001hellier}. Magnetic braking in some form \citep[e.g.][]{1983rappaport, 2022elbadry}, and gravitational wave radiation will drive the binary closer together, with accretion commencing in a few Gyr \citep[e.g.][]{2011knigge}. An extended discussion on the above points is presented in Appendix \ref{sec:extended_discussion}.

\section{Conclusions}
I have conducted the first phase-resolved spectroscopic observations of GLEAM-X J0704--37, an LPT pulsing on a 2.915-h period. The main conclusions are:
\begin{enumerate}
    \item RV shifts ($189\pm3\;\textrm{km s}^{-1}$) of the Na I doublet show that the binary orbital period (10496$\pm$5 s) agrees with the 10496.5522 s radio pulse period to within 5 (0.05) percent, fitting spectra from one (both) night(s) (Figure \ref{fig:rv}). 
    \item Radio pulses nearly coincide with the ascending node of the orbit, when the M dwarf is at maximal redshift and WD at maximal blueshift (Figure \ref{fig:rv}). This differs from what has been reported in ILT J1101+5521, where radio pulses arrive near binary conjunction.
    \item The average spectrum is well-fit by the sum of a $\approx$3,000 K, $\approx$0.14$M_\odot$ M dwarf and massive $\approx$7,300 K, $\approx$0.8--1.0$M_\odot$ white dwarf (Figure \ref{fig:best}; Table \ref{tab:params}).
    \item Weak H$\alpha$ emission is seen; some is attributed to the M dwarf, though some appears to be localized to a region outside the M dwarf that coincides with it in orbital phase, but orbits at a higher velocity of $\approx330\;\textrm{km s}^{-1}$ (Figure \ref{fig:map}).
    \item GLEAM-X J0704--37 ($\approx$400 pc) is nearly four times closer than previously thought ($\approx$1500 pc), suggesting that such systems may be relatively common.
    \item I propose that GLEAM-X J0704--37, along with ILT J1101+5521, are part of a class of LPTs that are associated with WD binary orbits, while other LPTs are likely associated with WD or NS spins.
\end{enumerate}

This work demonstrates the importance of multiwavelength follow-up in characterizing LPTs: optical spectroscopy has now revealed a class of LPTs associated with with WD + M dwarf close binaries \citep[this work;][]{2024discovery, 2024deruiter}, and X-ray pulsations along with a spatial association with a supernova remnant has tentatively connected another LPT to NS or WD spins \citep{2024wang,2024li}. Optical follow-up remains difficult, requiring the biggest telescopes in the world (10-m class) to adequately characterize optical counterparts. Ongoing searches for new LPTs may discover closer sources or reveal them to be an intrinsically distant, rare class of objects, in either case shining light on this exciting new phenomenon.

\begin{acknowledgements}
I thank Shri Kulkarni and Kareem El-Badry for a close reading of this paper and valuable feedback. \textrm{I also thank the referee for constructive input that improved the final manuscript.} I acknowledge support from an NSF Graduate Student Fellowship. I also thank Matthew Graham and Ilaria Caiazzo for obtaining an early spectrum that demonstrated the feasibility of this campaign and for assistance with observations, respectively. I thank the organizers and attendees of the \textit{XMM-Newton} ``The X-ray Mysteries of Neutron Stars and White Dwarfs'' Conference, including Natasha Hurley-Walker and Ingrid Pelisoli, for alerting me to the discovery of LPTs and to Myles Sherman, Casey Law, and the Caltech LPT Group for interesting discussions on related objects. I am grateful to the staff of Keck Observatory for their support in carrying out the observations presented here.\\

Some of the data presented herein were obtained at Keck Observatory, which is a private 501(c)3 non-profit organization operated as a scientific partnership among the California Institute of Technology, the University of California, and the National Aeronautics and Space Administration. The Observatory was made possible by the generous financial support of the W. M. Keck Foundation. I wish to recognize and acknowledge the very significant cultural role and reverence that the summit of Maunakea has always had within the Native Hawaiian community. We are most fortunate to have the opportunity to conduct observations from this mountain.
\end{acknowledgements}

%
%

\bibliographystyle{aa}
\bibliography{main}

\appendix

\section{Extended analysis: parameter constraints}

\subsection{Orbital solution}
\label{sec:appendix_orbit}
RVs are obtained by cross-correlating a 50\;\AA\; window around the Na I doublet with an M5 template from the BT-DUSTY library of theoretical spectral atmospheres \citep{2011allard}. The RVs are fit to Equation \ref{eq:rv} using data from both Night 1 and 2 simultaneously, assuming a circular orbit ($e=0$).
\begin{equation}
    \textrm{RV}_\textrm{MD} = K_\textrm{MD}\sin(2\pi(\phi-\phi_0)) + \gamma
    \label{eq:rv}
\end{equation}
``MD'' stands for M dwarf, RV is an individual RV, $K$ is the RV curve amplitude, $\phi$ is the orbital phase, $\phi_0$ sets the zero-point of the orbit (inferior conjunction), and $\gamma$ is the systemic velocity. To constrain parameter values,  a Markov Chain Monte Carlo (MCMC) parameter exploration \citep{1970mcmc} was performed using the \texttt{emcee} package \citep{2019emcee}; the ensemble sampler was run with twelve walkers for 10,000 runs, taking the first half as the burn-in period. Table \ref{tab:params} shows the credible intervals of all parameters, with the most relevant being $K_\textrm{MD} = 189\pm 3 \textrm{km s}^{-1}$. 

To calculate orbital phases, the period was first fixed to be equal to the radio period reported by \cite{2024discovery} of 10496.5522 s\footnote{Other periods were reported by \cite{2024discovery}, differing by at most 0.002 s. The current observing baseline yields a level of precision (5 s) which is consistent with all possible periods in \cite{2024discovery}.}. Phases were recalculated based on different candidate orbital periods in the range of 2.8 to 3 h and evaluated the fit to the RV solution derived above. This Monte Carlo trial constrained the  orbital period to be $P=10496 \pm 5$ s. However, since the total number of orbital cycles that transpired between Night 1 and 2 are unknown, performing the above analysis on the Night 2 data lone yields a more conservative error of $P=10500 \pm 500$ s. Both analyses are consistent with the radio period to within one sigma.

\subsection{Binary parameters}
\label{sec:appendix_binary}
To create an M dwarf model,  $T_\textrm{MD}$ is specified, interpolating through the BT-DUSTY library of theoretical stellar atmospheres \citep{2011allard}, assuming solar metallicity and $\log g = 5.0$. An inflation factor of 10\% is adopted for the radius of the M dwarf, as it has been shown that for binaries this close, fast rotation leads to increased magnetic activity and thus inflation \citep{2018parsons}. The stellar atmosphere is multiplied by $(R_\textrm{MD}/d)^2$ to obtain a flux as viewed from Earth. The same process is used for the WD, taking the theoretical (DA; H-rich) WD atmospheres of \citep{2010koester} and assuming $\log g = 8.0$.

A Markov Chain Monte Carlo (MCMC) parameter exploration was performed using the \texttt{emcee} package \citep{2019emcee} to explore all four simultaneously, adopting uniform priors on all parameters: $5000 < T_\textrm{WD}\;(\textrm{K}) < 12000 $, $2500 < T_\textrm{MD}\;(\textrm{K})< 4000$, $0.008 < R_\textrm{WD}\;(R_\odot)< 0.02$, $250 < d\;(\textrm{pc}) < 1000$. The MCMC sampler was run for 2000 steps, taking half as the burn-in period. The parameter exploration converged, with a Gelman-Rubin statistic \citep[$\hat{R}$;][]{1992gelman} of 1.08 averaged over all chains. \cite{1992gelman} argue that values close to 1, with $\hat{R}\lesssim1.1$ being a typical threshold, indicate convergence for typical multivariate distributions. The corner plot and  marginalized posterior distributions are shown in Appendix Figure \ref{fig:corner}. 

In Table \ref{tab:params}, the inferred binary parameters (median values and credible intervals: 16$^\textrm{th}$ and 84$^\textrm{th}$ percentiles of the posterior distributions) resulting from the MCMC analysis are shown. $M_\textrm{WD}$ is obtained from the mass-radius relation of \cite{2020bedard}; $M_\textrm{MD}$, from the M dwarf isochrones (assuming an age of 10 Gyr) of \cite{2015baraffe}, assuming solar metallicity. The binary separation, $a$, and inclination, $i$, are solved for using Kepler's laws, and the the M dwarf Roche lobe filling factor, $R_\textrm{MD}/R_L$, (i.e. how much of its Roche lobe is filled), is solved for using the \cite{1983eggleton} relation:
\begin{equation}
        \frac{R_L}{a} = \frac{0.49 q^{2/3}}{0.6 q^{2/3} + \ln\left(1 + q^{1/3}\right)}
        \label{ref:roche}
\end{equation}
where $q=M_\textrm{MD}/M_\textrm{WD}$ is the mass ratio between the two objects.

\subsection{Is GLEAM-X J0704--37 a magnetic WD or NS?}
\label{sec:appendix_wdns}

It is unlikely that the primary companion star to the M dwarf could be anything other than a WD. No main sequence star could fit in so compact an orbit. I speculate that in order to produce strong radio emission, the WD must be magnetic. Two known WD ``pulsars'' indeed host magnetic WDs in a 3--4 h orbits with an M dwarf \citep{2016marsh, 2023j1912} and are discussed further in Appendix Section \ref{sec:extended_discussion}. A NS is disfavored, since the blue excess (shortward of $\approx$6500 \AA) in the average optical spectrum would require a exceptionally young and nearby NS to be so optically bright \cite[e.g.][]{1997isolated_ns}. Being so nearby, such a NS would also be detectable in X-rays, which has been ruled out \citep{2024discovery}.

\section{Extended discussion: connections to cataclysmic variables and white dwarf pulsars}
\label{sec:extended_discussion}

\subsection{Comparison to ILT J1101+5521: cool and massive WDs}

The discovery of ILT J1101+5521 was reported by \cite{2024deruiter} as an LPT with a radio pulse period of 125.5195 min (2.0912 h). Optical spectroscopy of ILT J1101+5521 revealed an M dwarf with RV variability on a very similar period (127.4 min). It is likely that ILT J1101+5521 is indeed a detached WD + MD binary, as suggested by \cite{2024deruiter}, though absolute confirmation of this model requires spectroscopic observations over the entire period. Furthermore, both systems host rather cool, massive WDs: $\approx$7,300 K and $\approx$5,500 K; $\approx1.02M_\odot$ and $\approx0.77M_\odot$ for GLEAM-X J0704--37 and ILT J1101+5521, respectively. The WD in both systems appears to be more massive than the average isolated DA (hydrogen-atmosphere) and DB (helium-atmosphere) WDs: $0.6M_\odot$ and $0.7M_\odot$, respectively \citep{2007wd}. 

At the time of writing, the origin of radio pulses in LPTs has not been established, though a scenario for ILT J1101+5521 has been put forth \citep{2024qu}. In that model, electron cyclotron maser emission is responsible for the radio emission, requiring that the WD be strongly magnetic ($B\approx$ MG) and the M dwarf only mildly so ($B\approx$ kG). A similar situation could be the case for GLEAM-X J0704--37, though direct evidence for the magnetic nature of the WD (in either system) has not been established.

\subsection{Connection to CVs and the ``period gap''}

CVs are mass-transferring systems in which a WD accretes from a Roche lobe-filling donor, typically a late-type star \citep[e.g.][]{1995hellier, 2001hellier}. Magnetic CVs host a WD with a magnetic field strong enough so that the magnetosphere extends well past the WD surface and affects the accretion flow; recently, optical and X-ray surveys have revealed that 35--36\% of CVs are magnetic \cite{2020pala, 2024rodriguez_survey}. 

CVs typically have orbital periods in the range of $\approx$78 min--10 h, with an observed (though highly debated) ``period gap'' between 2.2--3.2 h \citep[e.g.][]{1983spruit, 2024schreiber}. It is believed that changes in the convective nature of the M dwarf donor star cause magnetic braking to become less efficient and accretion to temporarily stop \citep{1983spruit}. This has been supported by a higher observed rate of \textit{detached} WD + M dwarf binaries in this period range \citep{2016zorotovic}. The overlap of both GLEAM-X J0704--37 and ILT J1101+5521 with the CV period gap is exciting (Figure \ref{fig:two_types}), and may indicate that detached magnetic WD + M dwarf binaries in this orbital period range can exhibit a rich variety of radio activity. \textrm{However, it may very well be the case that GLEAM-X J0704--37 has \textit{never} undergone accretion, and is instead a post common envelope binary that will eventually become a CV (pre-CV). This is supported by the extremely low WD temperature, which may not have enough time to cool while crossing the gap ($\sim1$ Gyr), but would after emerging from common envelope ($\sim$ several Gyr). }

\subsection{LPTs, WD pulsars, and evolutionary links}

A connection between LPTs and so-called WD ``pulsars'' has been made by both \cite{2024discovery} and \cite{2024deruiter} regarding GLEAM-X J0704--37 and ILT J1101+5521. The two known WD pulsars, AR Sco and J191213.72--441045.1 (henceforth J1912), show pulsed emission at radio frequencies at 1.97 and 5.30 min, respectively \citep{2016marsh, 2023j1912}. Those periods are not the binary orbital period of the system, which are instead 3.56 and 4.03 h for AR Sco and J1912, respectively \citep{2016marsh, 2023j1912}. The emission mechanism in WD pulsars is unclear, with it currently being debated whether synchrotron or cyclotron dominates \citep{2017buckley, 2018stanway}. Unlike typical NS pulsars, it is thought that radio emission is generated through particle acceleration in the interaction between the WD and M dwarf magnetospheres \citep{2016marsh, 2023j1912}. Phenomenologically, there are two major differences between WD pulsars and long LPTs:
\begin{enumerate}
    \item \textit{Pulse periods in WD pulsars are shorter, and associated with WD spin periods}. In WD pulsars, radio pulses are associated with the spin-orbit beat between the WD spinning at min-long timescales and several hour-long orbits.
    \item \textit{Pulses in WD pulsars are 10--100 times fainter}. AR Sco and J1912 have peak radio luminosities of $\approx2\times10^{26} \textrm{erg s}^{-1}$ at $\approx$1 GHz. GLEAM-X J0704--37 has a radio luminosity of $\approx1.5\times10^{28} \textrm{erg s}^{-1}$ (at 1 GHz, adopting a distance of 400 pc in Equation 1 of \cite{2024discovery}), and ILT J1101+5521 has a typical peak radio luminosity of $2\times10^{27} \textrm{erg s}^{-1}$ with the highest peak reaching $1\times10^{28} \textrm{erg s}^{-1}$ (both at 100 MHz).
\end{enumerate}

I put forth the possibility that long LPTs represent a phase between WD pulsars and polars. Polars are the most abundant magnetic CVs \citep[e.g.][]{2020pala, 2024rodriguez_survey}, and typically have orbital periods below the period gap, in the 1.3--2.2 h regime \citep[e.g.][]{1995hellier, 2001hellier}. WD pulsars are detached, with the Roche lobe filling factor ($R_\textrm{MD}/R_L$) in AR Sco having been measured to be $\approx$0.8 \cite{2022pelisoli}. Since \citep{2022pelisoli} also detected that the WD in AR Sco is spinning down (to longer spin periods), WD pulsars may evolve into long LPTs. Magnetic braking in some form \citep[e.g.][]{1983rappaport,2011knigge, 2022elbadry} by the M dwarf will then lead it to fill its Roche lobe. At this point, accretion will commence and the system will be a polar Alternatively, accretion could commence before the WD fully spins down, which is consistent with the recent finding of a magnetic CV with a polar-like magnetic field, though with a WD spinning at 9.36 min \citep{2025rodriguez}. 

This is qualitatively similar to the picture put forth by \cite{2021schreiber}, further explored by \cite{2022ginzburg}, where WD pulsars are thought to evolve into polars. The WD magnetic field in this evolutionary model is thought to emerge as a result of crystallization in the WD core, which is more easily achieved in cool, massive WDs. Comparing to Figure 2 of \cite{2021schreiber}, it is clear that both the WD in in GLEAM-X J0704--37 and that in ILT J1101+5521 are cool and massive enough to be at least somewhat crystallized. Optical polarimetry of GLEAM-X J0704--37 should be undertaken in order to assess the magnetic nature of the WD. 

Finally, it could be that GLEAM-X J0704--37 and ILT J1101+5521 may be progenitors to magnetic CVs (post common envelope binaries) that have not yet experienced mass transfer. Their unusually cool WD temperatures are significantly cooler than those of WDs in a sample of detached WD + M dwarf binaries in this period range \citep[$T_\textrm{eff; WD}\gtrsim10,000$ K;][]{2016zorotovic}, and may favor this scenario.

\subsection{An emerging picture: two classes of LPTs}
\label{sec:two_classes}

\begin{figure*}
    \centering
    \includegraphics[width=0.9\textwidth]{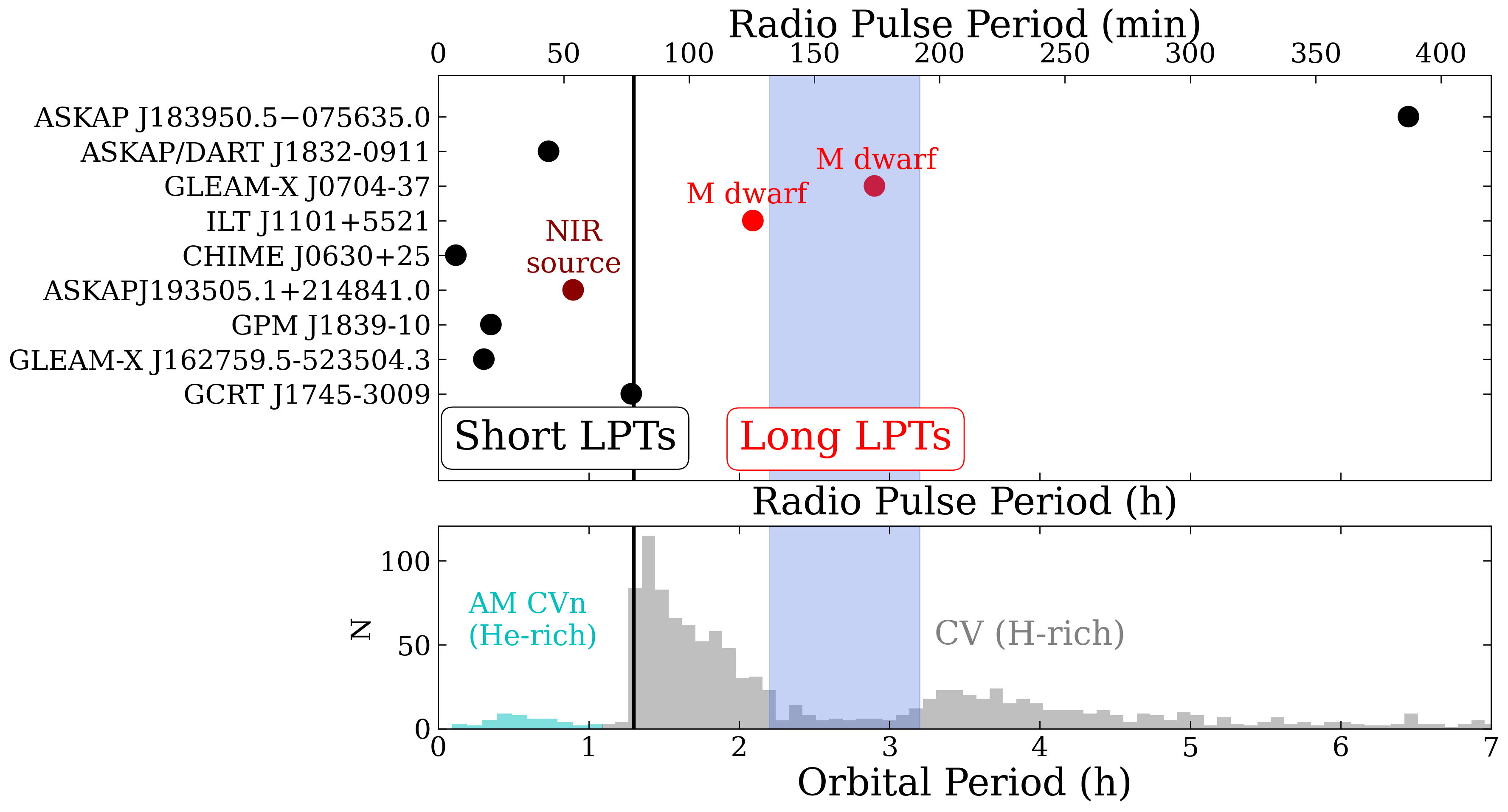}
    \caption{Two of the three ``long LPTs'' ($P\gtrsim$ 78 min) are associated with orbital periods of WD + MD binaries, while ``short LPTs'' ($P\lesssim$ 78 min) generally lack optical counterparts. The $\approx$78 min division (black line) corresponds to the orbital period minimum of any binary containing a Roche lobe-filling H-rich donor star, seen in the observed period distribution of CVs. The recent discovery of ASKAP J183950.5-075635.0 and its likely association with an NS spin suggests that ``long LPTs'' are not exclusively associated with WD orbits. Two of the three known ``long LPTs'' nearly coincide with the CV ``period gap'', where accretion is thought to temporarily shut off (blue band). }
    \label{fig:two_types}
\end{figure*}

I amplify the idea that there are two classes of LPTs: ``short'' ($P\lesssim78$ min) and ``long'' ($P\gtrsim78$ min)\footnote{Other works have also noted this distinction, though the relation to the CV period minimum has not been put forth.}. The dividing line between the two types in Figure \ref{fig:two_types} is the CV ``orbital period minimum'' of $\approx$78 min \citep[e.g.][]{2001hellier}. Samples of CVs and AM CVns are shown in Figure \ref{fig:two_types}, drawn from the catalogs of \cite{2003ritter} and \cite{2018ramsay}, respectively. However, there are two major caveats: 1) it is obvious that GCRT J1745-3009 is effectively on that line, and 2) the recent discovery of ASKAP J183950.5-075635.0 and its likely association with a NS spin reveals that ``long LPTs'' are not exclusively WD binaries. I primarily argue that ``long LPTs'' can be plausibly associated with orbital periods, but not ``short LPTs''. 

For \textit{any} mass-transferring system with a main-sequence star as the donor, $\approx$78 min is the minimum orbital period at which hydrogen burning on the donor star can be sustained \citep{1971faulkner, 1981paczynski, 2009gaensicke}. In other words, any binary system with an orbital period lower than this value cannot have a hydrogen-rich donor star\footnote{An interesting exception to this are hydrogen-rich donor stars with low metallicities, which can reach even shorter orbital periods \citep{1997stehle}.}. Due to angular momentum losses (from magnetic braking and gravitational wave radiation), long LPTs such as GLEAM-X J0704--37 will commence accretion near the orbital period minimum, and then ``bounce'' back to longer periods as CVs do \citep[e.g.][]{2011knigge}, meaning they will not overlap with the population of short LPTs.

Based on this, I suggest that in general, long LPTs are associated with WD binary orbital periods, and short LPTs with compact object (WD or NS) spin periods. Short LPTs may still be associated with orbital periods of binary systems, but unlikely with those in which the lower mass component is hydrogen rich\footnote{AM CVn binaries and ultracompact X-ray binaries (UCXBs) are examples of systems where donor stars dominated by helium and heavier elements orbit WDs and NSs, respectively, at periods below this minimum \cite[e.g.][]{2010nelemans}.}. The lack of optical counterparts in short LPTs would instead point to a spin period of an optically faint WD or NS. 

\section{Full spectra and RVs}

\begin{figure*}
    \centering
    \includegraphics[width=\textwidth]{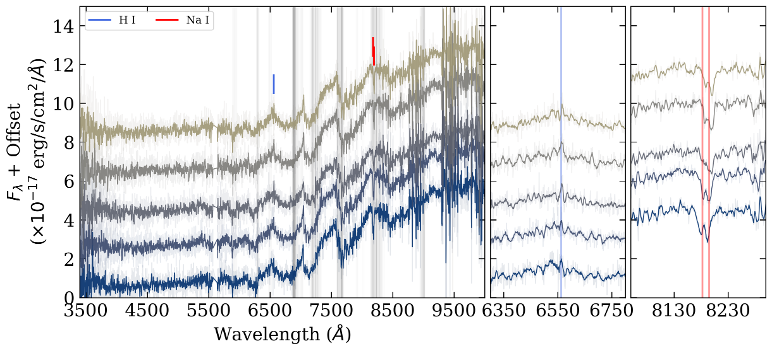}\\
    \includegraphics[width=\textwidth]{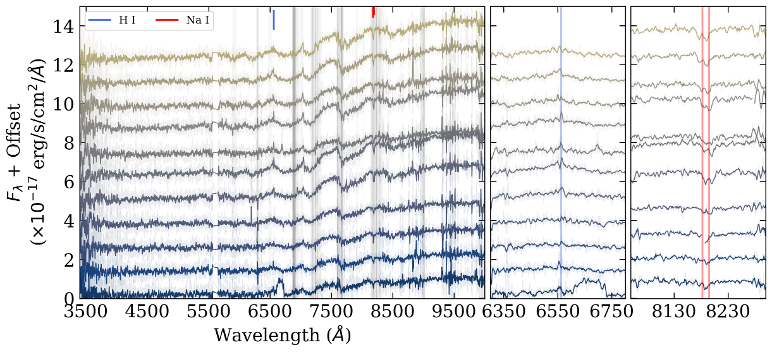}
    \caption{Keck I/LRIS 900-s exposures (top: Night 1, bottom: Night 2) of GLEAM-X J0704--37 reveal 2.9-h RV variability of an M dwarf star in the binary system. Gray bands indicate telluric features.}
    \label{fig:spectra}
\end{figure*}

Individual spectra from Night 1 and 2 are shown in Figure \ref{fig:spectra} with observing logs in Table \ref{tab:data}. All spectra are dominated by TiO molecular bandheads, characteristic of M dwarfs, as first revealed in the discovery paper by \cite{2024discovery}. I detect a clear Na I doublet (8183 and 8195 $\AA$) in absorption in all spectra, not seen in the original discovery paper due to wavelength constraints of the spectroscopic setup. I also detect weak H$\alpha$, which was not reported in the discovery paper, demonstrating the advatage of spectroscopy from 10-m class telescopes. The K I doublet (7665 and 7699\AA) is also seen in the spectrum, and yield an RV solution that is consistent with that of fitting the Na I doublet alone. However, the K I doublet is excluded from this analysis since it overlaps significantly with airmass-dependent telluric features that can alter the wavelength solution. No other obvious emission or absorption features are present.

\begin{figure}
    \centering
    \includegraphics[width=0.5\textwidth]{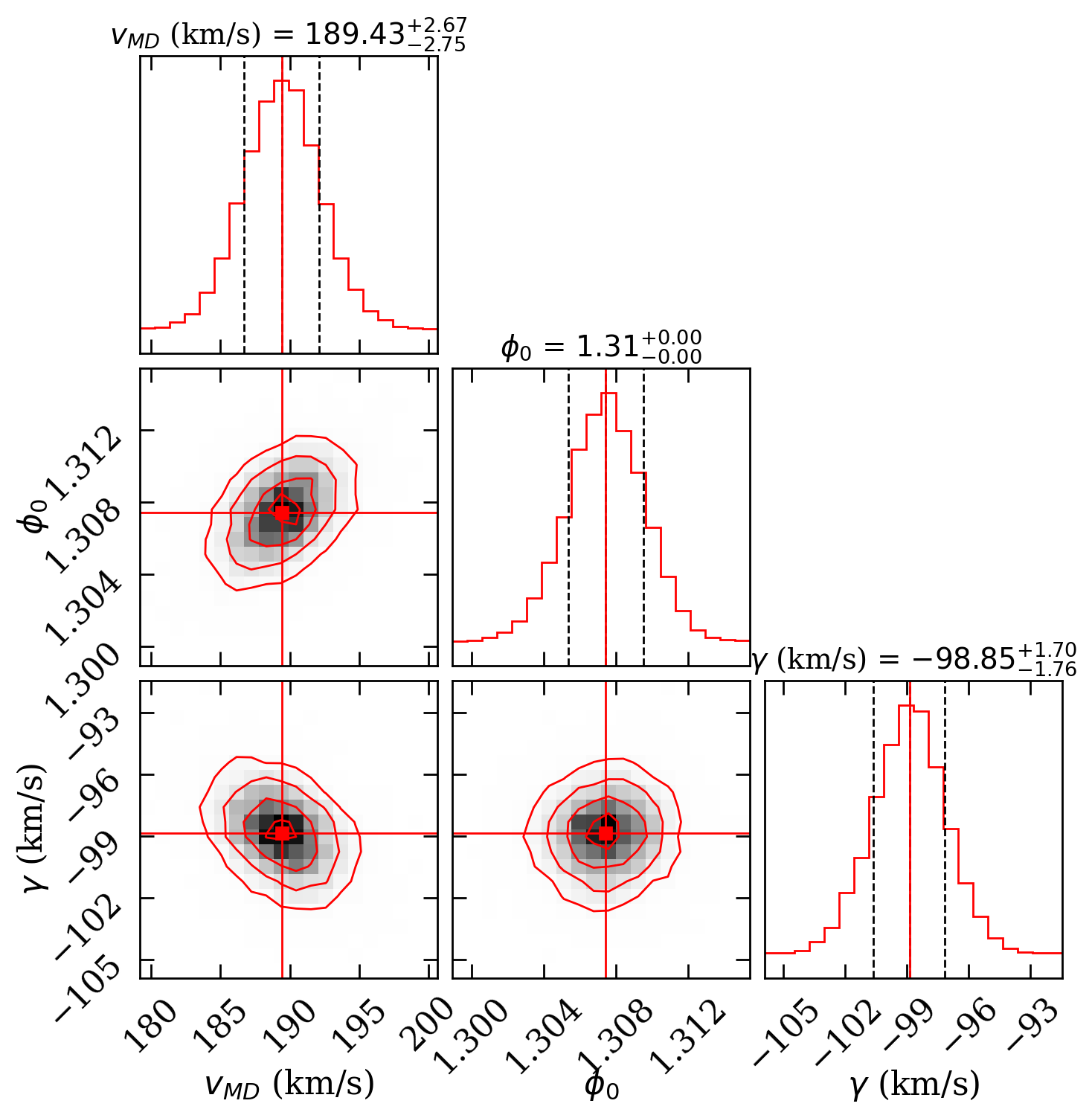}
    \caption{Corner plot of the MCMC analysis used to constrain RV parameters.}
    \label{fig:rv_corner}
\end{figure}

Even in the 1.3 h total observation from Night 1, radial velocity (RV) variations of the Na I doublet from the M dwarf are clearly seen. The measured RVs and ensuring MCMC parameter exploration of the RV model (Equation \ref{eq:rv}) are presented in Table \ref{tab:rv} and Figure \ref{fig:rv_corner}, respectively. The 3 h total observation from Night 2 shows the entire RV curve filled out, although by shallower Na I absorption lines, likely due to the presence of high clouds on that night. Some features in the red spectra of the second night could be attributed to cosmic rays or poor telluric subtraction, but are shown here for full transparency, namely the apparent emission feature at $\approx$6650--6670\AA\; in the first and seventh spectra of Night 2. Follow-up observations should confirm or refute the presence of such features.  

\begin{table}
    \centering
    \begin{tabular}{p{1.7cm}|p{1.32cm}|p{0.7cm}|p{2.2cm}|p{0.8cm}}
         Date& Exposures& Total Time (h) & Conditions&Airmass \\\hline
         Night 1: 08 Nov 2024&5$\times$900 s& 1.3&Light cirrus; 1.1" seeing&1.8--2.0\\
         Night 2: 30 Nov 2024&11$\times$900 s&3.0& Moderate cirrus; 0.9" seeing&1.8--2.5\\
    \end{tabular}
    \caption{Keck I/LRIS observation log of GLEAM-X J0704--37. The blue (red) side was binned at 2x2 (2x1) (spatial vs spectral axis). The 600/4000 grism was used on the blue side, and 400/8500 grating on the red side, leading to a resolution of approximately 1.2 and 1.1\AA\;, respectively. Total time includes read-out between consecutive exposures.}
    \label{tab:data}
\end{table}

\begin{table}[]
    \centering
    \begin{tabular}{l|c|c}
         MJD (BJD$_\textrm{TDB}$)& RV (km s$^{-1}$) & $\sigma_\textrm{RV}$ (km s$^{-1}$)\\\hline
         60622.579392 & -161 & 10\\
60622.590561 & -80 & 8\\
60622.601730 & 40 & 11\\
60622.612899 & 80 & 15\\
60622.624067 & 80 & 13\\
60644.432526 & -322 & 11\\
60644.443907 & -201 & 13\\
60644.455099 & - & -\\
60644.466292 & 0 & 3\\
60644.477483 & 80 & 3\\
60644.488663 & 0 & 10\\
60644.499879 & 40 & 30\\
60644.511071 & -40 & 5\\
60644.522264 & -161 & 5\\
60644.533444 & -241 & 6\\
60644.544625 & -282 & 6\\

    \end{tabular}
    \caption{RV measurements and errors of each exposure from Night 1 and Night 2. Times shown are the mid-exposure times (of 900s-long exposures in all cases), and have been barycentric-corrected.}
    \label{tab:rv}
\end{table}

\section{Origin of the H$\alpha$ emission, frame shifting,and cartoons of the geometry}
\label{sec:halpha}

Doppler tomography reveals the H$\alpha$ emission to be coincident with the orbital phase of the donor star, though with a higher RV. This means that the bulk of the H$\alpha$ emission originates from \textit{outside} the M dwarf. Figure \ref{fig:map}, however, \textit{is not} a spatial representation of the system --- rather, I explore two possible scenarios that could lead to the observed $\approx330\;\textrm{km s}^{-1}$ H$\alpha$ emission (which is higher than the $189\pm3\;\;\textrm{km s}^{-1}$ RV amplitude of the M dwarf):
\begin{enumerate}
    \item A compact emission region co-rotating with the M dwarf, but at higher orbital separation. 
    \item Emission from material falling onto the M dwarf.
\end{enumerate}
The geometry of both scenarios is visually outlined below. It is unlikely that the emission region would be located between the WD and M dwarf. The Doppler tomogram shows that the emission region is locked with the M dwarf at the same orbital phase, meaning that they have the same angular velocity, $\omega$. Since $v=r\omega$, where $r$ is the distance from the center of mass, this means that in order to see a higher velocity, the material would have to be located farther from the center of mass than the M dwarf. A way around this would be for the emission region to be in Keplerian orbit (i.e. not locked with the M dwarf) around the WD. However, this would mean that the material is localized to a small region (i.e. not a disk), and it is highly unlikely that we would see the orbital phase of the emission region coincide with that of the M dwarf in one, much less both observing runs. It could also be possible that the Na I absorption line could emerge from the side of the M dwarf facing the WD while H$\alpha$ emission originates from the center of mass of the M dwarf; however, this is unlikely given the large difference in velocities and previous findings in the CV literature \citep[e.g.][]{1988wade, 2011schwope}.

\begin{figure}
    \centering
    \includegraphics[width=0.5\textwidth]{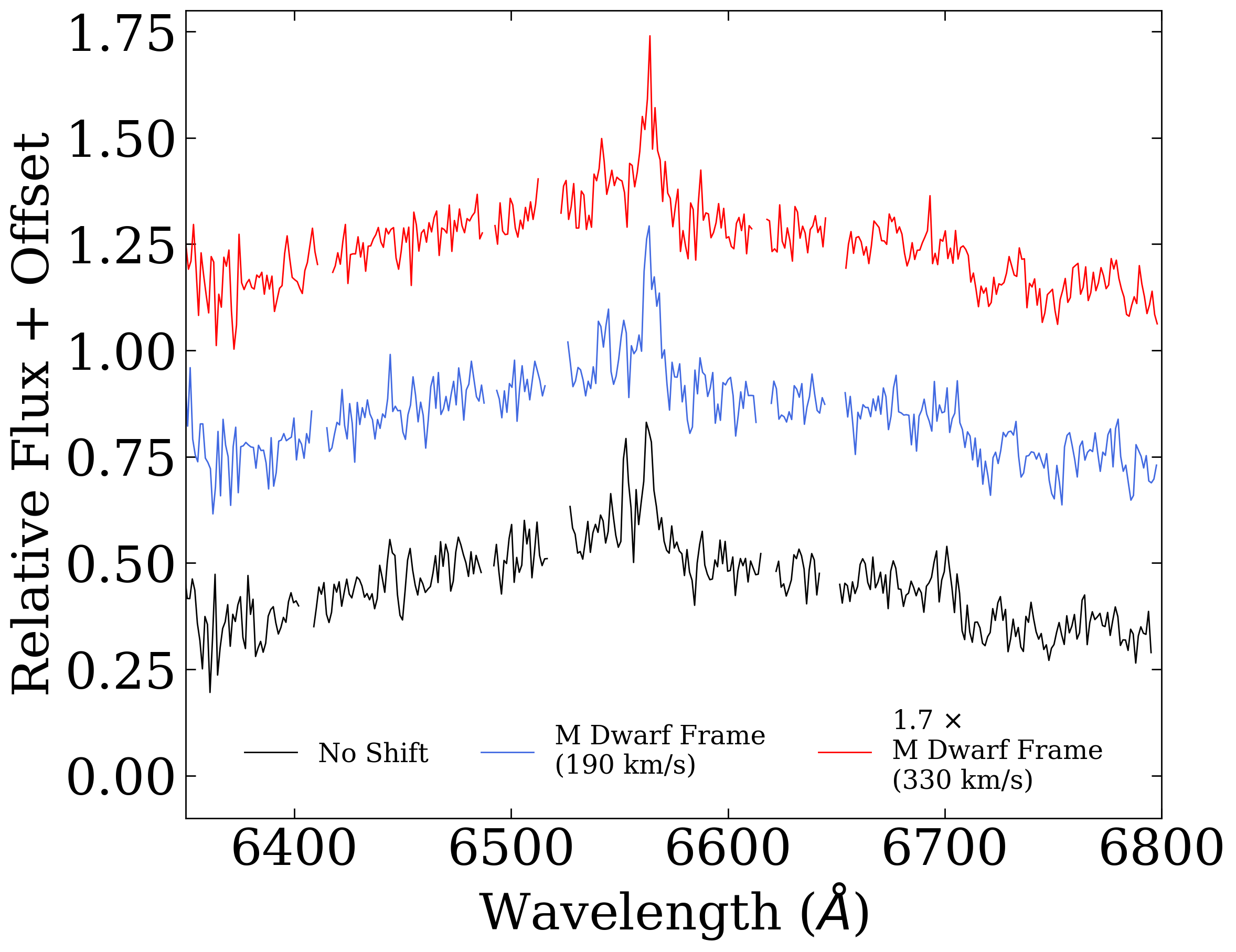}
    \caption{The average H$\alpha$ line profile after shifting into the reference frame of the binary (black), M dwarf (blue), and 1.7 times the velocity of the M dwarf (red). This demonstrates that H$\alpha$ is strongest and sharpest when shifting into the latter frame. This result only depends on the measured RV curve of the M dwarf from the Na I doublet, and is independent of any M dwarf parameter estimation.}
    \label{fig:ha_shifting}
\end{figure}

In Figure \ref{fig:ha_shifting}, I show the average spectrum around the H$\alpha$ line in the frame of the binary system (adding all spectra without any shifts; black), frame of the M dwarf (adding all spectra by shifting each one by the corresponding M dwarf RV; blue), and frame of the M dwarf but with an RV amplitude of $330 \textrm{km s}^{-1}$ (adding all spectra by shifting each one by 1.7 times the corresponding M dwarf RV; red). This shows that the strongest, sharpest emission is resolved when shifting into the $330 \textrm{km s}^{-1}$ frame, though signal-to-noise also builds up in the frame of the M dwarf. This exercise demonstrates that, independent of any model assumptions made, other than the measurement of the M dwarf RV curve with the Na I doublet, that H$\alpha$ emission comes from outside of the M dwarf.

\begin{figure*}
    \centering
    \includegraphics[width=\textwidth]{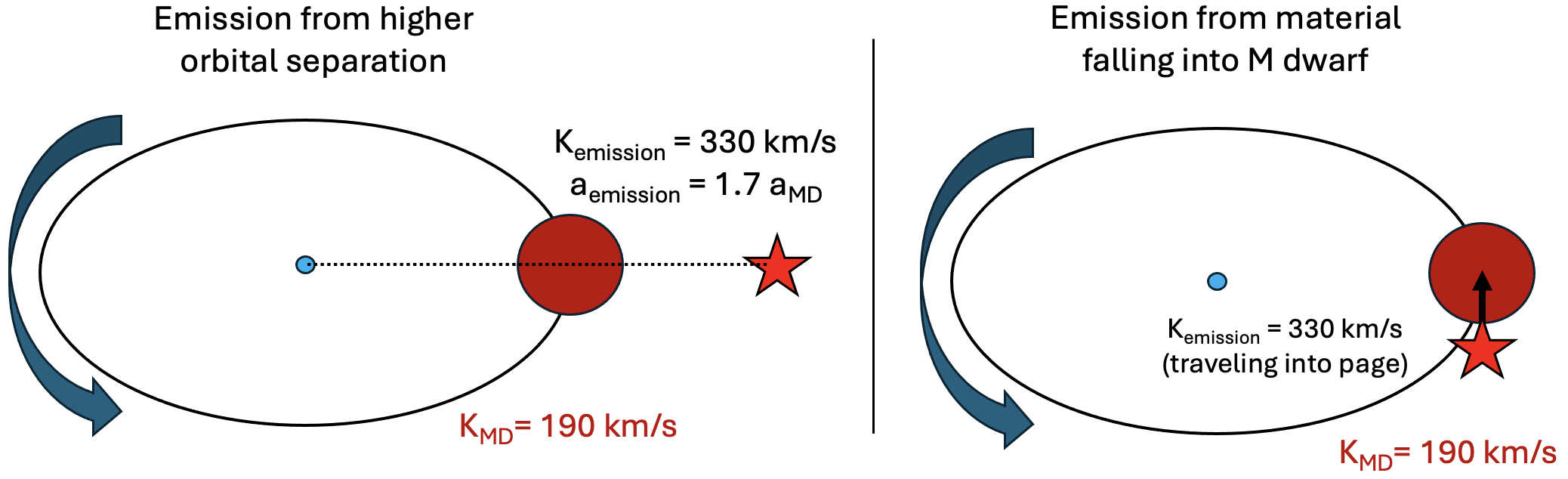}
    \caption{I propose two possible scenarios in which H$\alpha$ emission could be generated: 1) an emission region locked with M dwarf, co-orbiting the WD, though 1.7 times farther away from the WD; 2) an emission region originating from material falling into the WD, either through free-fall or by being magnetically channeled. Both diagrams are shown at the ascending node, where the radio pulses occur.}
    \label{fig:cartoon}
\end{figure*}

In Figure \ref{fig:cartoon}, I show cartoons of two possible scenarios that could lead to the observed H$\alpha$ emission. The two criteria to be met are: 1) the emission region must coincide with the orbital phase of the M dwarf and 2) must be seen to have an RV of 330 $\textrm{km s}^{-1}$, around 1.7 times that of the M dwarf. I speculate that the same phenomenon which powers the radio pulses also powers the H$\alpha$ emission, and I draw the cartoons in Figure \ref{fig:cartoon} in the orbital configuration (ascending node) where the radio pulses are observed (Figure \ref{fig:rv}). 

In the first scenario, material would be locked with the M dwarf, but located at a higher orbital separation, $a_\textrm{emission} = 1.7a_\textrm{MD}$. Taking $a=0.99R_\odot \Longrightarrow a_\textrm{MD} = 0.85R_\odot$, This corresponds to $a_\textrm{emission} = 1.5R_\odot$ and confirms that the emission must originate from \textit{outside} of the M dwarf. In the second scenario, the material would be falling in towards the M dwarf, potentially magnetically channeled. At that velocity, the free-fall height from the M dwarf center of mass would be $\sqrt{2GM_\textrm{MD}/h}$, corresponding to $h \approx 0.45R_\odot$, also well outside the surface of the M dwarf.

\subsection{connection to the 3.6-h post common envelope WD + M dwarf QS Vir}

\textrm{As a final note, the resemblance of this H$\alpha$ Doppler tomogram to at least one of QS Vir, a post common envelope WD + M dwarf in a 3.6-h binary only $\approx50$ pc away, is intriguing. Figure 7 of \cite{2010ribeiro} shows a similar picture to Figure \ref{fig:map} in this work --- some H$\alpha$ emission is associated with the M dwarf, yet a large fraction of it spills past the Roche lobe and originates from higher velocity. \cite{2010ribeiro} argued that this was evidence of magnetic confinement in a prominence-like magnetic loop. However, this is not the case in Figure 5 of \cite{2016parsons}, where spectra obtained at a later date revealed no H$\alpha$ emission at higher velocity, but instead some emission localized to different orbital phases. Recently, QS Vir was identified as a radio source in the VLA Sky Survey (VLASS) by \cite{2023ridder}. Follow-up of that source revealed a variable circularly polarized component in addition to a constant low-polarization component \citep{2023ridder}. However, it appears that QS Vir has not been observed for an entire orbital period in the radio, which is necessary to search for pulsations. Upcoming radio campaigns should target post common envelope-binaries to investigate the occurrence rate of LPTs among them. Curiously, a survey of WD candidates in a crossmatch between \textit{Gaia} and VLASS has revealed at most one candidate, suggesting that such systems may be rare, or radio-faint \citep{2024pelisoli}.}

\label{sec:h_alpha}

\section{MCMC corner plots}
\label{sec:mcmc_appendix}

\subsection{Prior information on distance and extinction}
\label{sec:distance}

GLEAM-X J0704--37 is located at Galactic coordinates of $\ell = 247.912298^\circ, b=-13.62183^\circ$. The low Galactic latitude corresponds to low, but non-negligible extinction: the 2D map of \cite{2011sfd} estimates a value of $E(B-V) = 0.25$ while the the 3D map of \cite{2022lallement} estimates a value of $A_\textrm{550 nm}= 0.821$ at its maximum reach of 800 pc, which corresponds to $E(B-V) \approx 0.28$. The map of \cite{2024edenhofer} is ultimately used, as described in the main text, which leads to $E(B-V) \approx 0.12$ . 

Notably, the distance to the system is poorly constrained \textit{a priori}. Its \textit{Gaia} DR3 (DR2) parallax of $-0.22\pm 0.98$ ($0.4\pm 1.1$) mas implies a distance of $3.9^{+3.0}_{-1.5}$ ($1.8^{+1.7}_{-0.9}$) kpc as estimated by \cite{2021bj} (\cite{2018bj}). \cite{2024discovery} report that electron density models of the Milky Way report lead to modest distance estimates of $0.4\pm0.1$ \citep{2017yao} and $1.8\pm0.5$ kpc \citep{2004cordes}. Since \textit{Gaia} does not detect a significant parallax, I favor an \textit{a priori} assumption that the distances implied by electron density models are favored, and expect the system to be located within 2 kpc, with the extinction to be at most $E(B-V) = 0.3$. 

\begin{figure*}
    \centering
    \includegraphics[width=0.5\textwidth]{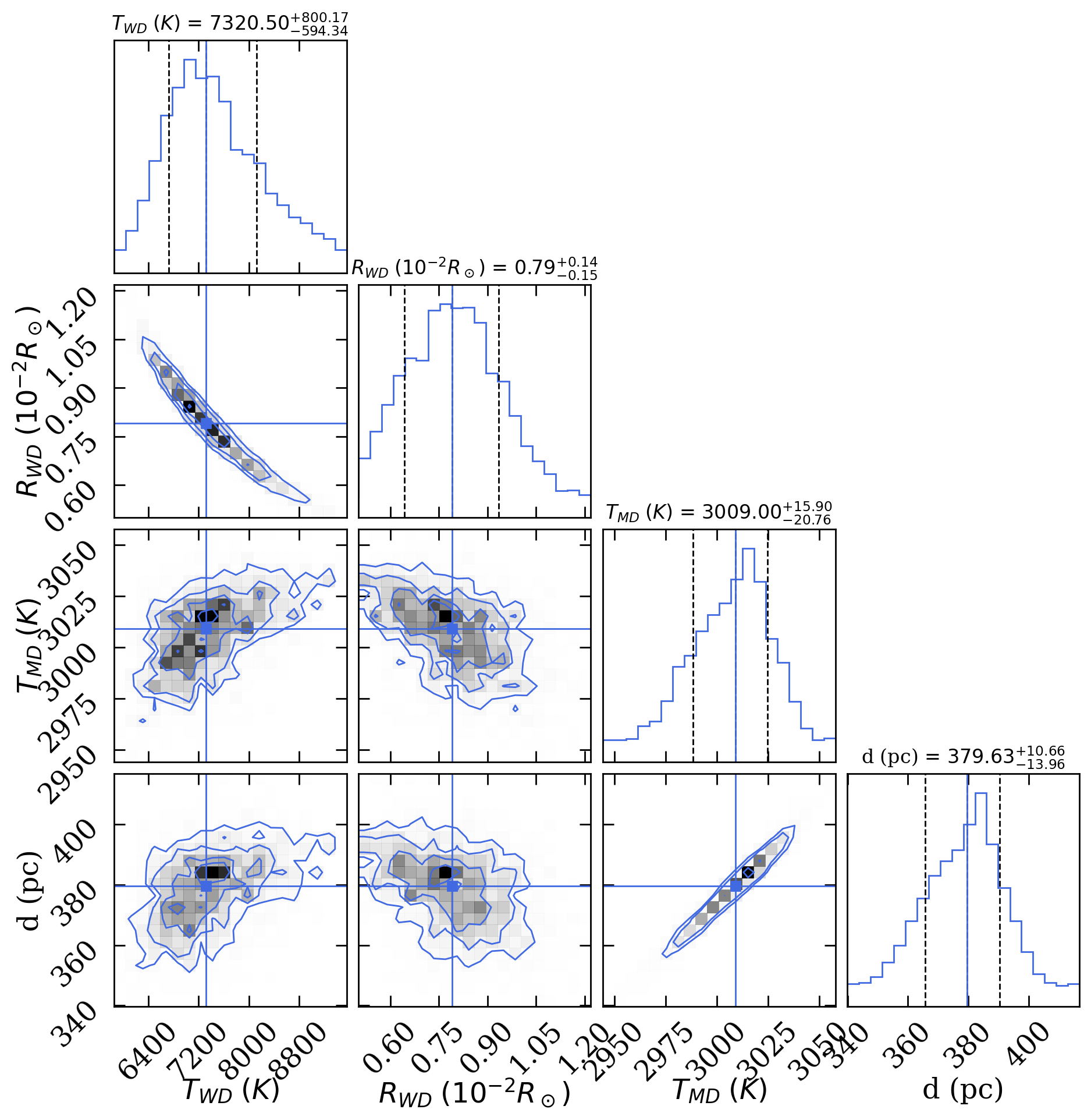}\includegraphics[width=0.5\textwidth]{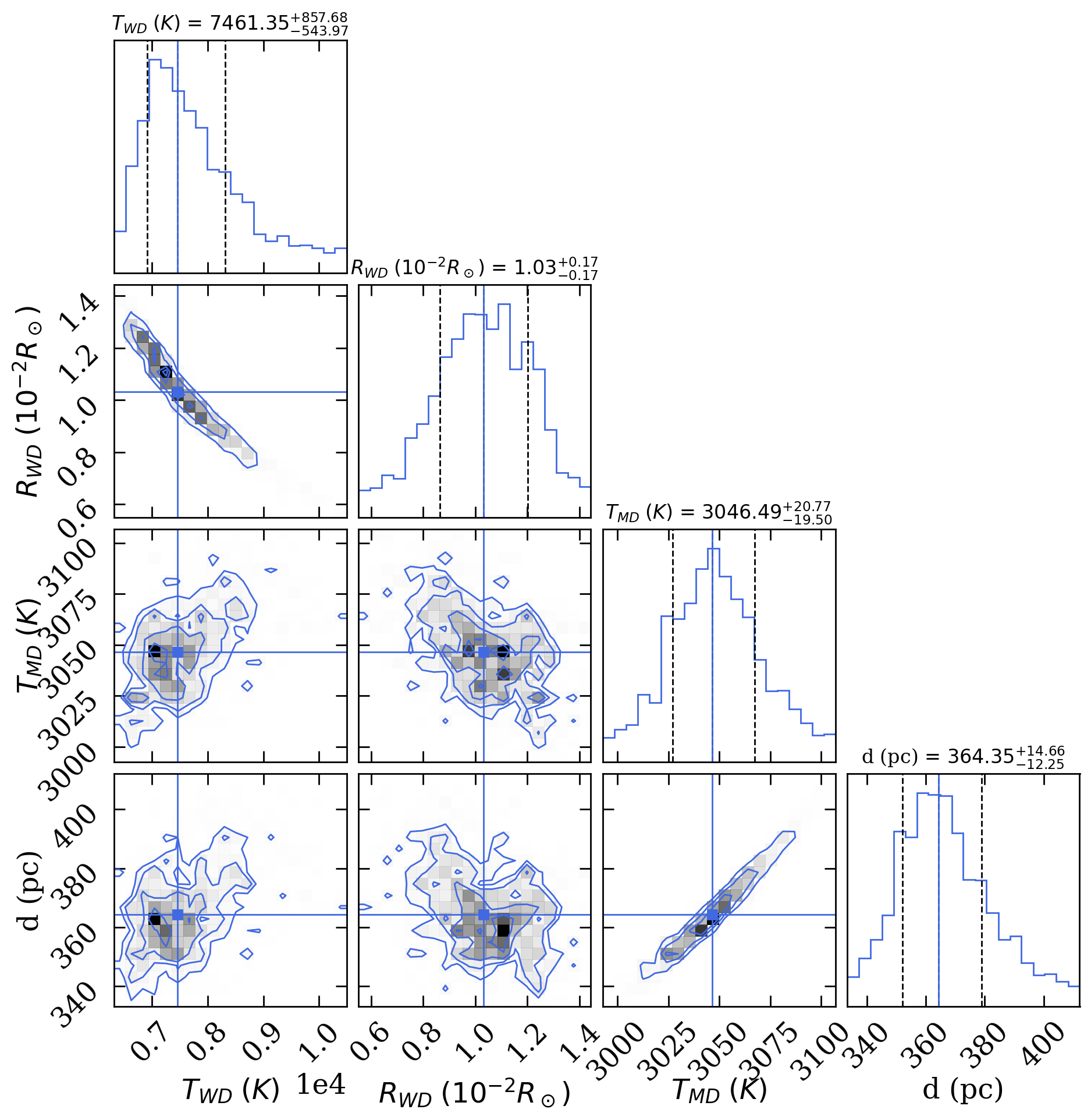}
    \caption{Corner plots resulting from the MCMC analysis. \textit{Left:} Letting $E(B-V)$ depend on the distance according to the 3D map of \cite{2024edenhofer}. \textit{Right:} Fixing $E(B-V)=0.3$, which is likely the highest possible value. This leads to a lower inferred WD mass, though which still exceeds that of the mean single WD in the field \cite[0.6 $M_\odot$ for a DA WD;][]{2007wd}. }
    \label{fig:corner}
\end{figure*}

In addition to the MCMC analysis described in the main text, I ran another routine identical to the one adopted in the main text, but fixing $E(B-V) = 0.3$. This was done in order to produce the lowest possible mass WD (since higher extinction values lead to higher WD radii). The table of parameters and ensuing corner plot are shown in Table \ref{tab:params_0.3} and Figure \ref{fig:corner}, respectively. Even assuming extreme values of extinction, it appears that the WD in the system is massive, taking on a value of $0.82 M_\odot$ with $E(B-V) = 0.3$. 

\begin{table}[]
    \centering
    \begin{tabular}{l|c}
        Parameter & Value \\
         \hline
         $E(B-V)$ & 0.3 (fixed)\\
         $T_\textrm{WD}$ (K) & $7460^{+860}_{-540}$\\
         $R_\textrm{WD}$ ($R_\odot$) & $0.0103^{+0.0017}_{-0.0017}$\\
         $T_\textrm{MD}$ (K) & $3050 \pm 20$\\
         $R_\textrm{MD}$ ($R_\odot$) & $0.172^{+0.005}_{-0.004}$\\
         $d$ (pc) & $364^{+15}_{-12}$\\
         \hline
         $M_\textrm{WD}$ ($M_\odot$) & $0.82^{+0.14}_{-0.12}$\\
         $M_\textrm{MD}$ ($M_\odot$) & $0.143 \pm 0.005$\\
         $a$ ($R_\odot$) & $1.01^{+0.05}_{-0.04}$\\
         $q$ & $0.174^{+0.030}_{-0.026}$\\
         $R_\textrm{MD}/R_L$ & $0.698 \pm 0.001$\\
         $i$ ($^\circ$) & $32^{+2}_{-3}$\\
         \label{tab:params_0.3}
    \end{tabular}
    \caption{Same as Table \ref{tab:params}, but fixing $E(B-V)=0.3$. The parameter that is most strongly affected is the inferred WD mass, reducing the median value from 1.02 to 0.82$M_\odot$. Since higher values of extinction are unlikely, it appears that the WD must be above the mean WD mass of single WDs \citep[0.6$M_\odot$;][]{2007wd}.}
    \label{tab:table_a}
\end{table}

\end{document}